# A machine learning approach to predict *L*-edge x-ray absorption spectra of light transition metal ion compounds

Johann Lüder[1,2,*]

[1]*Department of Materials and Optoelectronic Science, National Sun Yat-Sen University, 70 Lien-Hai Rd., Kaohsiung 80424, Taiwan, R.O.C.*

[2]*Center of Crystal Research, National Sun Yat-Sen University, 70 Lien-Hai Rd., Kaohsiung 80424, Taiwan, R.O.C.*

(Dated: July 28, 2021)

## Abstract

Simulation of spectra of x-ray absorption spectroscopy (XAS) at the *L*-edge is a well-established and reliable computational tool that, in combination with experimental measurements, reveals details about the local electronic structure and the chemical environment of transition metal ions (TMI). For light TMIs, model 2*p* XAS model Hamiltonian approaches (MHA) are fast and accurate in computing spectra. Here, an alternative method using artificial neural networks (ANN) is employed. The new approach predicts x-ray absorption spectra at the *L*-edge of light transition metals ions for elements from Ti to Ni, including different oxidation states, from sets of physical input parameters. Computational performance and accuracy were optimized through a systematic exploration of ANN design and compared to MHA simulations. While the MHA remains more accurate due to the statistical nature of the developed machine learning method, the ANN-based prediction of 2*p* XAS spectra achieved good accuracy of more than 99%, in addition to a noticeable advantage in computational speed. This makes the method intrinsically more suitable for computational high-throughput setups as well as in support of data-intense experimental workflows. However, the extrapolation weakness of artificial neural networks remained for most cases and displayed a strong dependence on the TMI. For one TMI, an extrapolation accuracy of 78% was maintained.

# Introduction

X-ray absorption spectroscopy (XAS) at the *L*-edge is a well-known technique to investigate and understand the nature of 3*d* states in, for instance, transition metal compounds, the ion's local chemical environment and their resulting physical and chemical properties.[1–3] Because XAS is an element and state selective non-destructive analysis method, this technique is routinely used for characterizations of materials and to elucidate mechanisms at the atomic scale, for instance, inducing changes of atomic and electronic structure [4–8]. Through its wide range of possible applications, it is used for investigations in many technological relevant domains in which 3*d* electrons dominate atomistic mechanisms and materials' functionality such as in catalysis, chemical energy storage and in the design of magnetic devices [8–12].

In a transition metal, an electron from a semi-core 2*p* energy level is excited into an unoccupied 3*d* state in the 2*p* XAS process. In light transition metals, the complex nature of this excitation due to spin-orbit coupling, core-hole valence interaction and the rich multiplet features, can provide challenges in simulation of 2*p* XAS spectra by mean-field approaches. The foundation of accurate simulation of 2*p* XAS spectra was laid several decades ago [13,14] and was based on atomic multiplet calculations, which allowed efficient computations. Spectra of light TMI in many compounds were well reproduced, including the non-trivial branching ratio of $L_3$ and $L_2$ features. Also effects caused by changes in the chemical surrounding, which crystal fields can describe, are accounted for, and their characteristic spectral features are reproducible[15]. Later, charge transfer effects were also considered and the model was extended to ligand field effects [16,17]. This model approach, which reduces the XAS process on a subspace of electrons including the TMI's valence *d*-, semicore *p*- and *p*-electrons of the surrounding atoms, has become a frequently used and reliable scientific tool through, e.g., the CTM4XAS interface [18] and is widely used.

Nowadays, more computational methods to simulate 2*p* XAS spectra are available. Most approaches are either based on model Hamiltonians, including charge transfer models, [2,19,20] or on first-principles approaches mostly either using some form of time-dependent density functional theory [21–23], a variation of the Bethe-Salpeter equation to a subspace of 2*p* and 3*d* states [24–28], configuration interaction methods also working with subspaces, or related approaches [29]. Both model and first-principles approaches have advantages: speed and ease of computational demands for the former and independence of parametrization, and a subsequent transferability to untested systems for the latter. Also, combinations of first-principles methods with a model Hamiltonian ansatz were proposed [4,30,31]. However, there are not yet frequently employed in practice while further efforts are devoted to continuous developments of this research direction [32]. A comprehensive review of theoretical 2*p* XAS methods applied to transition metals is given, for instance, in Ref. [33].

In recent years, machine learning (ML) lead to significant technological advances. Today ML impacts various research fields, including medicine, physics, chemistry, and materials science [34–39]. However, many ML techniques have been developed for tasks in computer science such as image and voice recognition. Because those fields are very different from the original applications, the capability of ML techniques and implementations in other areas needs to be explored.

In materials-related research, reported ML achievements become more frequent as computational resources, data and efficient implementations become widely available. In particular artificial neural networks (ANNs) are now applied to various problems [40,41]. For instance, ANNs 'solve' Schrödinger's equation [41,42] (or the rather similar Kohn-Sham equations [43,44]), accelerate method developments enabling faster computations and more accurate molecular dynamics simulations [45], provide predictions of new materials and their properties for applications in pharmaceuticals, solar and energy storage, catalysis, etc. [46–54], or they improve, accelerate and advance computational material science [55–62]. While their application in some areas is only recently feasible due to computational advances and large datasets, some areas such as potential energy surfaces ANNs are explored for several decades [63–65]. Many different machine learning techniques can be explored for their beneficial use in experimental and theoretical approaches. Also, combinations of experimental and theoretical methods can be employed by ML tools. For instance, Schwenker *et al.* linked experimental and theoretical imaging techniques for interface and defect mapping [66]. More detailed reviews can be found in, for example, Refs. [40,41,67].

In x-ray absorption spectroscopy, ML techniques were investigated by several authors for applications related to the K-edge. In the K-edge process, a core electron from a 1s state is excited by a photon. Carbone *et al.*[68] presented a machine learning approach to predict K-edge spectra of molecules containing N and C atoms. Also, a classifier approach that can detect local coordination environment from K-edge spectra reaching an accuracy of 87 % was presented [69]. Better accuracy of 97 % was achieved for K-edge XANES in a two-layer neural network to classify the local chemical structure of $Li_3FeO_{3.5}$ [70]. A bit later, Rankine *et al.*,[71] also showed that ANN can predict XANES K-edge spectra [71]. Other types of spectroscopy allow spectra prediction by neural networks, too. Gosh *et al.* [72] showed that neural networks can achieve 97 % accuracy in predicting excited state spectra for molecules. The work by Timoshenko *et al.* focused on determining structure factors in extended x-ray absorption fine structure signals taken at the K-edge, and was applied to nanoclusters [73,74] and bimetallic nanoclusters [75] in addition to their recent review on related work [76]. Marella *et al.* extended the idea to an ANN that can determine the partial coordination numbers in bimetal (PdAu) nanoclusters [77]. An unsupervised approach employing an autoencoder neural network architecture was proposed as well [78] to gain information about important structural features from XANES. Of course, other machine learning techniques besides artificial neural networks are in use for application in spectroscopy. For instance, Torrisi *et al.* [79] employed the random forest techniques on K-edge spectra, which allowed them to predict Bader charges on atoms and nearest neighbor distances and maintain interpretability of the machine learning model enabling to analyze the impact of data pre-processing on interpretability.

While machine learning methods applied to K-edge x-ray absorption spectroscopy are advancing, less work was devoted to XAS at the *L*-edge. At the *L*-edge, electrons from the 2*p* state are excited by photons. In contrast to K-edge excitations, the confinement of 2*p* orbitals and their overall with other states significantly impacts the physical processes. Then a multitude of possible transitions can lead to relatively complex spectra, especially in light transition metals. In fact, XAS at the *L*-edge has certain advantages when investigating TMI and provides access to element and state selective information.

Recently, a proposed machine learning approach allowed inversion of 2*p* XAS spectra at the *L*-edge obtaining physical properties such as the *d*-level splitting, among others as part of the parametrization of a 2*p* XAS model Hamiltonian [80]. An artificial neural network-based implementation was explored for this purpose. A 2*p* XAS model Hamiltonian simulated spectra for the training of the model. The trained models yielded a parametrization of a 2*p* XAS model Hamiltonian from an experimental spectrum. Even for experimental spectra containing small amounts of linear background signals and white noise [80], the model performed well. This is the so-called inverse problem. In most applications, ANNs are used to extract information in a similar matter from data which leads to some form of dimensional reduction from input with $N$ features to output with $M < N$ predictions such as image recognition [81–83], parameter extractions [80,84], prediction of properties in materials [85–87].

In contrast to the previous work from Ref. [80], here a spectrum containing a much larger number of channels than the number of input parameters shall be generated, i.e., the direct problem will be solved by ANN. The presented ANN approach, which is illustrated in Figure 1, aims to generate information of a spectrum with $C$ channels as output from $P$ parameters where generally $C \gg P$.

Here, ANNs are presented to predict x-ray absorption spectra at the *L*-edge for light transition metals ions for elements from Ti to Ni, i.e. $Z = 22 \ldots 28$, including different oxidation states and symmetries of atomic arrangements of nearest-neighbor atoms from small sets of physical input parameters. A total of 22 TMIs having between 2 and 8 electrons were chosen. Each spectrum has an energy window of 30 eV that starts 5 eV below the ions' $L_3$ edge. Additionally, a systematic exploration of ANN design helped to increase the accuracy of ANNs. The analysis of ANN performance/accuracy was supported by statistical means as well as real-world applications. The presented method has potential for applications in high-throughput tools such as computational workflows for materials discovery and characterization, and automized evaluations through feature fitting in experimental results.

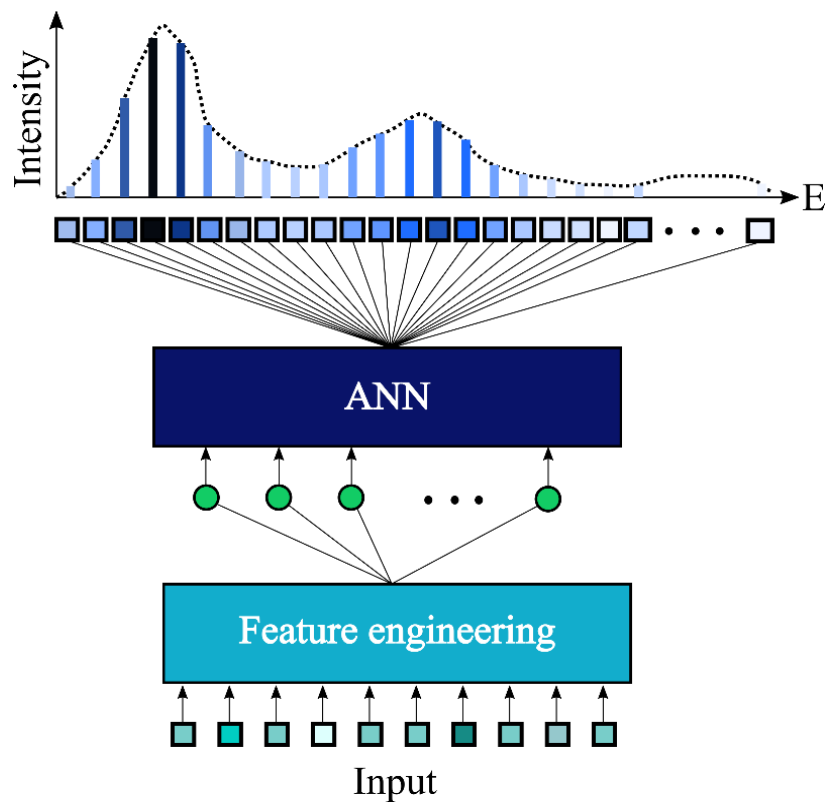

*Figure 1. Principal workflow of the proposed ML method to generate 2p XAS spectra from a set of input parameters.*

## Methods

### Model Hamiltonian-based database

Here, a database of 360000 spectra for each of the tested transition metal ions (i.e. a total of 7.92M spectra), which also includes different oxidation states, in a $D_{4h}$ symmetric crystal field was generated with the Quanty code [31,88,89]. The simulation of $2p$ XAS spectra is based on the model Hamiltonian approach described below. Within the approach, parameters express physical conditions and properties. Only a brief discussion on the $2p$ XAS model Hamiltonian approach (MHA) is given.

*Table 1. Information about the parameter ranges to generate databases for different TMI.*

|      | $D_q$/eV | $D_s$/eV | $D_t$/eV | $S_1$ | $S_2$ | $T$/K | $\sigma_G$/eV | $\Gamma_1$/eV | $\Gamma_2$/eV |
|------|----------|----------|----------|-------|-------|-------|---------------|---------------|---------------|
| Min. | 0.0      | -1.4     | -0.8     | 0.6   | 0.6   | 1     | 0.05          | 0.3           | 0.3           |
| Max. | 0.2      | 1.4      | 0.8      | 1.0   | 1.0   | 1000  | 0.20          | 1.0           | 1.2           |

For the $2p$ XAS process, the initial state model Hamiltonian is given in Eq. (1)

$$\hat{H} = \sum_{i,j} \epsilon_{d_{i,j}} \hat{d}_i^\dagger \hat{d}_j + \sum_{i,j} \epsilon_{p_{i,j}} \hat{p}_i^\dagger \hat{p}_j + \zeta_d \sum_{i,j} \langle d_i | \vec{l} \cdot \vec{s} | d_j \rangle \hat{d}_i^\dagger \hat{d}_j + \zeta_p \sum_{i,j} \langle p_i | \vec{l} \cdot \vec{s} | p_j \rangle \hat{p}_i^\dagger \hat{p}_j$$
$$+ \sum_{i,j,k,l} U_{ijkl}^{dd} \hat{d}_i^\dagger \hat{d}_j^\dagger \hat{d}_l \hat{d}_k + \sum_{i,j,k,l} U_{ijkl}^{dp} \hat{d}_i^\dagger \hat{p}_j^\dagger \hat{p}_l \hat{d}_k$$

(1)

where $\epsilon_{d_{i,j}}$ and $\epsilon_{p_{i,j}}$ describe the relative energy positions of TMI valence d- and semi-core p-states, respectively. The Hubbard U-terms account for strong coulombic interactions between electrons, in particular between d-d ($U^{dd}_{ijkl}$) and d-p ($U^{dp}_{ijkl}$) electrons. The splitting of state due to spin-orbit coupling ($\vec{s} \cdot \vec{l}$) in valence ($\zeta_d$) and core ($\zeta_p$) is considered as well. Note that $\hat{d}^{\dagger}_i$ and $\hat{p}^{\dagger}_i$ create d- and p-electrons while $\hat{d}_j$ and $\hat{p}_j$ annihilate them. So-called Slater Condon integrals, for which precomputed values [90] were obtained by single atom calculations in the framework of Hartree-Fock theory, express a multipole expansion of Coulombic and exchange interaction between electrons. The mentioned Hubbard U-term represents a part of the monopole term in an expansion, and it must, in general, be empirically determined. To account for effects by solid-state phase and missing correlation effects in the Hartree-Fock theory, screening factors $S_1$ and $S_2$ are introduced for Coulomb and exchange interactions, respectively. This aims to relate single atoms conditions to those of atoms within a solid.

Within a single transition metal atom, d-states are degenerated. The local chemical surrounding of an ion within a solid or molecule can lift this degeneracy and can be described of a crystal field. Ballhausen parameters $D_q$, $D_s$ and $D_t$ can express resulting relative energies of d-states as

$$\epsilon_{d_{a_{1g}}} = 6D_q - 2D_s - 6D_t$$

$$\epsilon_{d_{b_{1g}}} = 6D_q + 2D_s - D_t$$

$$\epsilon_{d_{b_{2g}}} = -4D_q + 2D_s - D_t$$

$$\epsilon_{d_{e_g}} = -4D_q - D_s + 4D_t$$

(2)

where $d_{a_{1g}}$, $d_{b_{1g}}$, $d_{b_{2g}}$ and $d_{e_g}$ refer to $d_{x^2}$, $d_{x^2-y^2}$, $d_{xy}$ and $d_{xz}, d_{yz}$ states, respectively.

The solutions of the eigenequation $H|\psi_i\rangle = E_i|\psi_i\rangle$ provide access to eigenstates $|\psi_i\rangle$ and energies of each configuration $E_i$ as well as a simulated $2p$ XAS intensity $I(\omega)$ of a spectrum through Eq. (3)

$$I(\omega) = -\frac{1}{Z}\sum_i \Im\left(\left\langle\psi_i\left|\widehat{D}^\dagger \frac{1}{\omega - \widehat{H} + E_i + \frac{i\Gamma}{2}}\widehat{D}\right|\psi_i\right\rangle\right)\exp(-\beta E_i)$$

(3)

In Eq. (3), the dipole operators $\widehat{D}$ obtain the transition intensities from initial to a final state, in which a $2p$ electron was promoted into an unoccupied d-state, at excitation frequency $\omega$. The transition intensities are summed up and weighted by $\beta$, the inverse of the temperature, and the partition function $Z$. Also, the imaginary shift $\Gamma$ from the real axis results in a Lorentzian broadening [90,91], which is usually modeled with a stepwise-defined frequency dependent function because the lifetimes of core-holes in $p_{\frac{1}{2}}$ and $p_{\frac{3}{2}}$ states differ. After lifetime broadening, a Gaussian broadening with an empirically determined broadening factor $\sigma_G$ can be applied to account for experimental convolution. The selection of parameter ranges, from which randomly selected sets of parameters were used to generate the databases of spectra, follows the previous work in Ref. [80]. In Ref. [80], databases for $Ni^{2+}$, $Co^{2+}$, $Fe^{2+}$ and $Mn^{2+}$ were computed. Here, it also includes other oxidation states and more transition metal elements. An overview of the used parameters and parameter ranges is given in Table 1.

The simulation of $2p$ XAS with Eq. (1)-(3) depends on sets of physical input parameters which form the descriptor or input vector for the ANN. Note that all input and output data were box normalized, i.e., the smallest and largest values of the parameters and in a spectrum were scaled to 0 and 1, respectively. A computed spectrum on 501 channels represents the output or feature vector. Each dataset of input and output vectors, i.e. respective sets of parameters and spectra, for each metal ion can be split into training, evaluation and test datasets. Training data and evaluation data are used to perform the weight optimization, i.e. ANN training, and monitor the fitting process of the ANN, respectively. Evaluation datasets can also be used to optimize hyperparameters (those being used besides the weights of the ANN) such as learning rates, momentum factors in optimization algorithms, etc. In contrast, test datasets are not employed for the training or optimization of the ANN but serve to determine the performance of the trained and optimized ANN.

## Artificial Neural Networks

Artificial Neural Networks are complex constructs of simple building blocks, so-called neurons represented by mathematical functions, organized into so-called layers which are used to pass information from an input to an output layer [92]. Between input and output layers are the hidden layers. Each neuron in one layer is connected to all the neurons in the other layer in dense or fully-connected feed-forward neural network architectures. As information is passed to neurons and processed by simple mathematical functions, the weights on each neuron can be optimized during training to match training data. This rather adaptable design of ANN allows them to learn abstract relations between input and output. It is known that ANN can represent intricate and unknown relations in data, in principle, to arbitrary accuracy [93].

While ANNs were initially investigated to model processes in brains, they found application in text, voice and image recognition [94–96] later and are now entering many yet ANN unexplored areas of science and technology, which is supported by modern implementation frameworks such as Keras [97]. The Keras application programming interface is also being used in this study. To address the question of which type and ANN design are most suitable for the task of simulating 2$p$ XAS spectra, four different ANN architecture types based on dense, convolutional, transpose convolutional and locally connected neural networks are tested. In dense neural networks (DNN), as mentioned above, all neurons in a layer are connected to each of the neurons in the next layer. A general expression of DNN that passes information from the input vector $x$ to the output $y$ is given by

$$y_l = f_o \sum_{k_N=0}^{D_N} \omega_{j,k_N}^{(N)} f_{N-1,k_{N-1}} \left( \sum_{k_{N-1}=0}^{D_{N-1}} \omega_{k_N,k_{N-1}}^{(N-1)} f_{N-2,k_{N-2}} \times \cdots \left( \sum_{i=0}^{D_i} \omega_{k_1,i}^{(1)} x_i \right) \right) \tag{4}$$

where activation functions $f$ processing input (i) to output component (o), and each of the $N$ layers having $D_j$ neurons and weights $\omega_j$ (biases are omitted). The architecture of a DNN is illustrated in Figure 2a where the green neurons represent input neurons, cyan ones belong to a pre-processing layer (dense as well) and the yellow neurons are neurons of the fully connected network to which information is passed to, i.e., either an output layer or the next hidden layer. The architecture of ANN can be more complex than the comparatively unstructured DNN. In DNN, all neurons are connected but not all connections might be relevant for the model. Then, the associated processing units do not lead to benefits. At the same time, computational constraints for very deep and wide DNN can be lifted by more adaptable ANN types such as convolutional neural network. Convolutional neural networks (CNN) connections between neurons are somewhat more selective and structured by applying kernels of width $w_f$ across neurons in a layer [98]. In CNNs, the convolution in one dimension at component $m$ can be expressed as

$$\omega[m] \otimes x[m] = \sum_{i=-h}^{h} \omega[i] \cdot x[m-i] \tag{5}$$

A kernel refers to the subarrays used to scan over all input and filter size refers to the number of neurons in the layer that processes the information from the kernel. A general architecture of a CNN is shown in Figure 2b where purple neurons represent the filter and the boxes around cyan neurons the kernels. Note that in practice, the strides (i.e. the shift between kernels) can be as small as one, whereas the figure shows a stride of two.

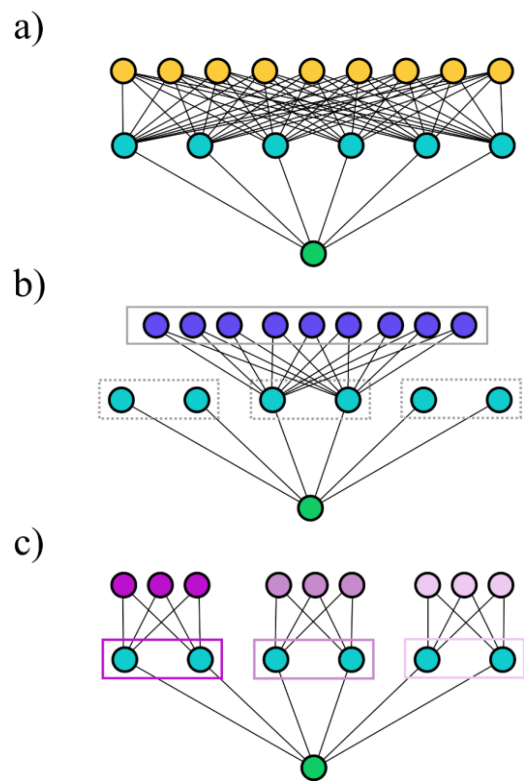

*Figure 2. Overview of schematic representations of a) dense, b) convolutional and c) locally connected neural networks where information is passed from an input neuron (green) to a fully-connected layer (cyan) which then connects to the specialized architectures in DNN (yellow), CNN (blue) and LCNN (magenta). Note that groups are marked with boxes for which the weights in b) are optimized for all groups while each group/filter in c) obtains its own set of weights.*

Locally connected neural networks (LCNN), illustrated in Figure 2c, work similar to CNN but the weights in the filter are individually optimized for each kernel, while in CNN the weights in the filter are averaged for all kernels. Lastly, the transpose convolutional layers (TNN) perform the inverse operation of a CNN.

Here, nine parameters are passed to the ANN to generate a 2p XAS spectrum including L$_3$ and L$_2$ peaks of at least 500 channels. The input parameters include three Ballhausen parameters $D_q$, $D_s$, and $D_t$, screening factors $S_1$ and $S_2$, lifetime factors $\Gamma_1$ and $\Gamma_2$, experimental broadening and external temperature $\sigma$ and $T$, respectively. An intermediate feature engineering step (FES) was used before passing the physical parameters to the ANNs, indicated in Figure 1. In FES, new features in products of the form $g^n h^m \ldots i^p$ were formed where $g, h \ldots, i$ are values of normalized inputs and possible combinations of $n, m, \ldots, p \in \mathbb{Z}$ with $n, m, \ldots, p \geq 0$ and $n + m + \cdots p \leq M$; $M$ is denoting the largest exponent. In this way, a $M = 3$ generated a total of 252 input features.

The difference between predicted and actual spectrum $j$ at neuron/channel $i$ is given as $\Delta_{i,j} = y_{p_j,i} - y_{a_j,i}$. Following Ref [80], the spectrum error (SE) can evaluate the performance of the trained ANN. SE is the mean average percentage error (MAPE) between the spectrum computed by the model Hamiltonian (including lifetime factors and experimental convolutions) and the ANN predicted spectrum, i.e., $SE_j = \sum_i \frac{\Delta_{i,j}}{\|y_{p_j,i} + y_{a_j,i}\|}$. Here, the channel resolved SE ($SE_c = 1/N \sum_{j=0}^{N} \Delta_{i,j}$) is introduced as well as the mean of all test spectra ($\mu_i = 1/N \sum_j^N y_{a_j,i}$) and their standard deviation, i.e. $\sigma_i = \sqrt{\sum_j^N (y_{a_j,i} - \mu_i)}$.

With the box normalization, SE values of less than 3% generally give excellent visual agreement, while up to 7% is still in good agreement. At SE larger than 10%, differences become noticeable in spectra shapes (this included peak positions and peak features such as peak width, shoulders and tails). Hence, a SE of 10% is chosen here to be the threshold deciding if a spectrum is correctly predicted, i.e. a *match*, or not, i.e. an *outlier*. Of course, this choice is not absolute and some form of quantitative influence remains while the qualitative picture should be valid over other reasonable choices of this accuracy threshold.

# Results & Discussion

## Optimization and training

The performance of ANN depends on efficient training and the architecture of ANNs. Here, four different classes of architecture were optimized and tested to predict 2$p$ XAS spectra for $Ni^{2+}$. Here, ANN classes differ by the prevalent employed type of hidden layers. The different architecture classes were chosen to be dense, transposed convolutional, convolutional, and locally connected layers in ANN; then ANN are referred to as DNN, TCNN, CNN and LCNN, respectively.

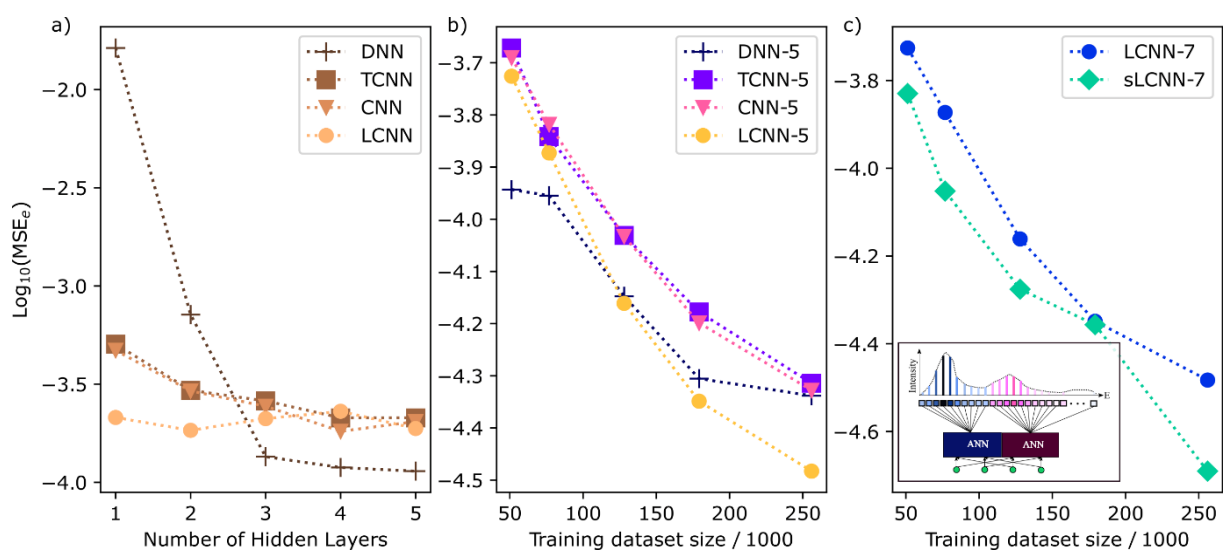

*Figure 3. Comparison of the impact of ANN architectures (number of hidden layers, type of hidden layers) and training dataset size on MSE for the evaluation datasets.*

For each architecture class, the ANNs were constructed by the corresponding layer type in hidden layers. For all ANNs, input and output layers were of the dense type. Additionally, two dense layers between the input layer and the hidden characteristic layers were introduced. This allows control of the networks' complexity as well as introduces additional processing units. The dense layer connected to the input layer had 260 neurons and the other had 160 neurons. Furthermore, each architecture class was optimized with respect to the number of hidden characteristic layers, i.e., the depth of a ANN. In general, deep networks perform better than wide and flat networks [80,98]. LCNN, CNN and TNN had 165 neurons as filters and a kernel size of 25 in each hidden characteristic layer. DNNs had 1000 neurons in the characteristic hidden layers. All but the output layer, for which a linear function was selected, used the ReLU (Rectified Linear Unit) as an activation function. Other activation functions, including swish [99], PReLU (Parametric Rectified Linear Unit) and hyperbolic tangent, were tested with no significant nor consistent improvement of ANN performance. Weights in ANN were optimized by the root mean square propagation algorithm.

Figure 3a shows the evolution of the validation mean squared error (MSE) for four different ANN architecture classes with different amounts of hidden layers from one to five characteristic hidden layers, each training on 32000 spectra with a training-evaluation split of 8:2. DNN exhibits the largest improvement from 0.016 to less than 4.6e-5 with an increasing number of hidden layers. LCNN, CNN and TNN also show an improvement, but it is less pronounced with MSE values between 5.1e-4 and 3.3e-5. While there is a separation in MSE for LCNN and CNN/TNN with only a few hidden layers, they approach similar values for five hidden layers.

Figure 3b reveals that the number of training data affects the ANNs' accuracy. Data for training/evaluation was increased from 32k to 320k in multiples of 32k and a training-evaluation split of 8:2 was applied as well. Differences in the relative ordering of performance of ANN architecture classes, each ANN with five hidden characteristic layers, are significant. At the smallest amount of training points, DNN outperforms the other ANN while LCNN and

DNN reach similar MSE for 160k data points. Thereafter, LCNN outperforms the other ANNs. Both, CNN and TNN, show similar MSE outcomes.

A further modification of the ANN, which is motivated by the physics of the 2$p$ XAS process, was introduced. In fact, the processes and effects that lead to the different shapes of and the non-trivial branching ratio between $L_3$ and $L_2$ peaks in 2$p$ XAS spectra can in principle be modeled individually, which was accounted for by a split in the ANN into two separated sub-ANN, one for $L_3$ and the other for $L_2$, as indicated in the inset in Figure 3c. The same figure gives the resulting increase of accuracy seen by a decrease of MSE from 3.3e-5 to 2.0e-5 for seven-layer LCNN trained on 320k data points.

*Table 2. Training results of sLCNN-7 after 80 epochs were reported as MSE [MAE].*

|     | Ti | V | Cr | Mn | Fe | Co | Ni |
|-----|----|----|----|----|----|----|----|
| 2+ | 4.60e-05 [0.0034] | 4.33e-05 [0.0034] | 3.64e-05 [0.0032] | 3.23e-05 [0.0030] | 2.34e-05 [0.0025] | 2.19e-05 [0.0023] | 7.61e-06 [0.0011] |
| 3+ | - | 4.89e-05 [0.0037] | 3.66e-05 [0.0033] | 3.65e-05 [0.0034] | 3.67e-05 [0.0033] | 2.09e-05 [0.0025] | 2.02e-5 [0.0023] |
| 4+ | - | - | 5.93e-05 [0.0044] | 5.39e-05 [0.0043] | 4.19e-05 [0.0038] | 3.831e-05 [0.0034] | 2.87e-05 [0.0029] |
| 5+ | - | - | - | 6.74e-05 [0.0049] | 5.25e-05 [0.0044] | 4.21e-05 [0.0039] | - |
| 6+ | - | - | - | - | 6.95e-05 [0.0051] | - | - |

The ANN split architecture of LCNN was selected with seven hidden characteristic layers (further decreasing the MSE to 1.76e-5 from five characteristic layers) for the remaining discussion, from here on referred to sLCNN-7. Lastly, the impact of *a priori* imposed FES should be analyzed. A comparison of sLCNN-7 without FES shows a MAE 3.38e-5 which is 66% larger than with FES.

Careful tuning of ANN design, architecture and prepossessing steps can help to decrease evaluation MSE by 62% for ANNs trained on large datasets. Table 2 and Table 3 report the mean absolute error (MAE) and the MSE for training and validation, respectively, obtained after 80 epochs with sLCNN-7 for all tested elements and their different oxidation states. For the training data, MAEs are between 0.1e-3 and 0.4e-3 and MSE are less than 0.5e-4. Notably, error values for training are smaller than for validation by a factor of about 2 in MSE. Overfitting was not observed, and the sLCNN-7 ANN design worked rather well for all tested ions. The data show two trends: i) with increasing oxidation state MSE/MAE worsen and ii) MSE/MAE increase for elements having a single atom configuration closer to a half-filled or empty d-shell. The latter point reflects the fact that for single atoms the number of possible electronic configurations increases with a half-filled shell.

## Testing and performance analysis

Testing the performance of ANN should go beyond reported averages of error values such as MAE and MSE because they cannot provide a detailed picture of type of error sources. In this regard, they cannot answer how critical the largest errors are and the overall distribution of errors. In particular, the largest errors may lead to catastrophic failures of machine learning tools for practical applications and might stay undetected through an overall well performing ANN. In addition, a sufficiently large amount of test cases should be employed to capture the statistical nature of the applied method [93]. Here, 10000 sets of test parameters and corresponding test spectra for each element and its oxidation state were selected. Test spectra were not used for training or optimization of ANNs. The SE was evaluated between test spectra

(those directly obtained from the model Hamiltonian approach) and the predicted spectra, i.e. those obtained from a trained ANN.

Table 3. Validation results of sLCNN-7 after 80 epochs were reported as MSE [MAE].

|    | Ti       | V        | Cr       | Mn       | Fe       | Co       | Ni       |
|----|----------|----------|----------|----------|----------|----------|----------|
| 2+ | 1.12e-04 [0.0046] | 1.02e-04 [0.0046] | 9.13e-05 [0.0043] | 3.24e-05 [0.0030] | 5.90-e05 [0.0033] | 4.62e-05 [0.0029] | 1.76e-05 [0.0014] |
| 3+ | -        | 1.40e-04 [0.0038] | 1.10e-04 [0.0050] | 9.42e-05 [0.0046] | 8.98e-05 [0.0044] | 6.05e-05 [0.0035] | 4.69e-5 [0.003] |
| 4+ | -        | -        | 1.65e-04 [0.0064] | 1.39e-04 [0.0061] | 1.13e-04 [0.0053] | 1.08e-04 [0.0048] | 7.96e-05 [0.0041] |
| 5+ | -        | -        | -        | 1.98e-04 [0.0073] | 1.46e-04 [0.0065] | 1.17e-04 [0.0056] | -        |
| 6+ | -        | -        | -        | -        | 2.06e-04 [0.0077] | -        | -        |

Figure 4 summarizes the SE of all ion's test data by dividing them into different intervals for sLCNN-7. This allows comparing the reliance and performance of the ANN for different ions. Besides the SE scoring categorizations mentioned above, additional intervals between 3% and 10% SE were introduced. The additional intervals give a better resolution of performance differences. In total, seven intervals divide the SEs. The first interval ($I_1$) represents all SE of less than 3 %, previously identified as excellent reproduction of spectra [80]. On the other end of the scale are SEs of more than 10 % ($I_7$), that, in general, show large discrepancies (those spectra are here classed outliers) between ANN generated and the reference spectrum. Note that simple shifts of spectra or scaling effects can be their cause as well. Between SEs of 3 % and 10 %, intervals are split by 1 % steps, i.e. $I_2$ to $I_6$. Note that still good accuracy was identified for SE of less than 7 %, which is the union of $I_1$ to $I_5$.

For all ions, the latter union contains 99 % of all test cases, while at least 88 % reach excellent accuracy with SE < 3 % ($I_1$). Most ions except $Ni^{4+}$, $Mn^{5+}$ and $Fe^{6+}$ reach SE of less than 3% for more than 91% of all test data. Note that these three cases belong to ions with a half-filled d-shell in the final state. The three mentioned ions represent cases with the largest amount of combinatorial arrangement of five d-electrons in ten d-states as given in their final state of $2p$ XAS excitation. Additionally, there is a weak correlation of decreasing $I_1$ with an increasing oxidation state for each element.

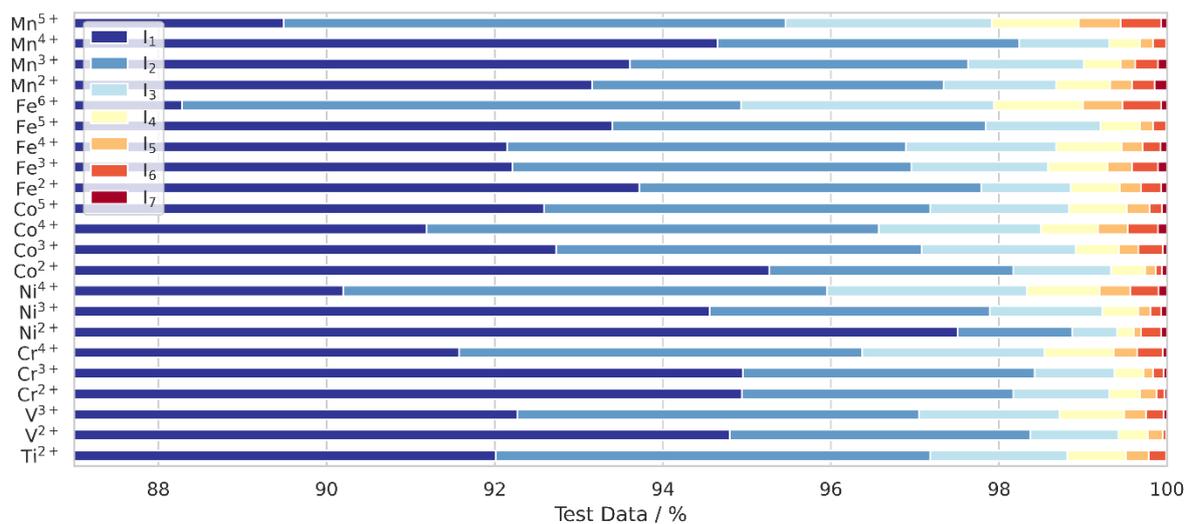

Figure 4. SE of test data in intervals $I_1$=[0,0.03[, $I_2$=[0.03,0.04[, $I_3$=[0.04,0.05[, $I_4$=[0.05,0.06[, $I_5$=[0.06,0.07[, $I_6$=[0.07,0.10[ and $I_7$=[0.10,1.00]. The latter interval marks so-called outliers that have strong differences to reference spectra.

It should be noted that the here claimed accuracies and their categorizations depend on their measures. For large SE, there might be no similarity between spectra or one of the $L_3$ or $L_2$ features is not well reproduced. For smaller SEs, several types of misfits can occur between reference and AI-predicted spectra. They include, for instance, single peaks over/underestimates or the already mentioned small shifts. Categorizations into different intervals of MAPE are one out of many choices. For example, comparing relative peak positions of the *M* most dominating features in spectra is possible as well. For 2*p* XAS, however, broad tail regions can be caused by cascades of small intensity transitions, or small details in features between large peaks, as they can occur experimentally in, for example, FeO [100] or Co-phthalocyanine [101] (or the well-known temperature sensitivity of $Co^{2+}$ 2*p* XAS spectra), challenge simple evaluation schemes of this kind. In addition, an extensive inclusion of spectral regions with no intensity (such as regions far from $L_3$ and $L_2$ features) generally results in misleadingly improved MSE/MAE because the ANN can reproduce zero intensities for those regions without difficulty and, in this way, artificially lower the overall average of errors. Therefore, a direct comparison to other studies' reported accuracies obtained with other measures or energy windows might be misleading.

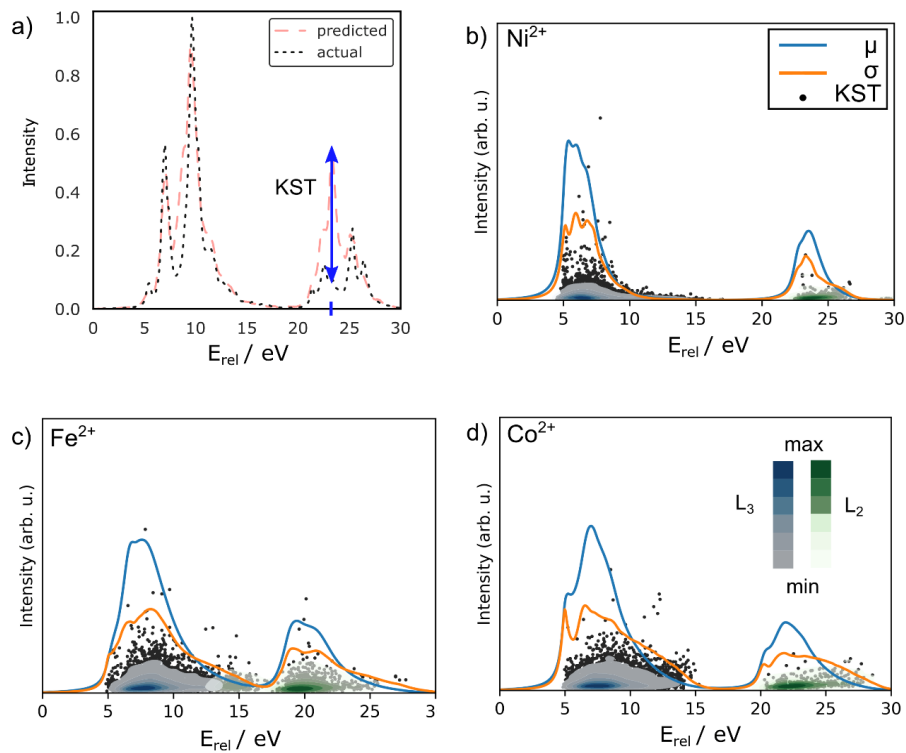

*Figure 5. a) Schematic of KST/MEPD determination in which the blue intersection with the energy axis marks the position and the double arrow the value of the KST; b), c) and d) MEPD, μ and σ of the test database for $Ni^{2+}$, $Fe^{2+}$ and $Co^{2+}$, respectively. The k-means determined clusters of the MEPD and the resulting kernel distribution estimate are superimposed as isocontour plots in with 6 segments.*

SE, however, averages the total intensity of misfits in spectra. As such SE could undervalue individual peak misfits. Because the presence or absence of a single peak in a spectrum can be indicative of electronic structure effects, it is important to distinguish in more detail the causes of SEs. Hence, a statistical analysis method that is more sensitive to individual peak misfits is needed.

Here, the Kolmogorov-Smirnov test (KST), a statistical measure to compare normal distributions, is adapted for this purpose. In KST, the similarity between two distributions is compared by determining the largest difference and its position. This idea can be applied to the difference in predicted and actual spectra. Then, the largest difference corresponds to the largest peak misfit ($\Delta_{i,j}$), that is either an underestimation or overestimation of peak intensity, and its energy. Only absolute values are considered here. This is illustrated in Figure 5a. The data distribution resulting of KST applied to all test spectra will be called maximal erratic peak intensity distribution (MEPD). For each ion's MEPD, the energy positions are passed to a k-mean clustering algorithm to determine if the KST position belongs to $L_3$ or $L_2$.

In an unsupervised k-mean cluster algorithm, a hyperparameter determining the number of clusters must be selected. Two clusters are assumed: one for $L_3$ and the other for peaks belonging to $L_2$. As shown later, this assumption is justified because the channel, i.e. neuron, resolved errors are proportional to the branching ratio between $L_3$ and $L_2$, and KST distributions follow $2p$ XAS intensities. Because 10000 KSTs are in one MEPD, density distributions of the $L_3$ and $L_2$ clusters in the MEPD are obtained with kernel distribution estimates (KDE) which serve as visualization aid. Three examples of the resulting MEPDs and the clustering of $L_3$-$L_2$ peaks, including the KDEs can be seen in Figure 5b) to d) for selected elements. The figure also superimposes the average ($\mu_i$) and standard deviation ($\sigma_i$) of the test spectra. Note that all test spectra had normalized intensities between 1 (maximum) and 0 (minimum).

From the examples of $Ni^{2+}$, $Fe^{2+}$, and $Co^{2+}$ shown in Figure 5 it is evident that most KST misfits are indeed centered around the main peaks of $L_3$ and $L_2$. For these ions and the others (not shown), the distribution of KST follows somehow $\mu$ and $\sigma$. The KDEs also show that most KSTs have small $\Delta_{i,j}$. Although the vast majority of max. peak intensity errors are small or negligible, often around 100-200 data points have noticeable peak intensity mismatches, i.e., the value of the KST. The KDE of MEPD are in good agreement with the channel resolved mean values of the test spectra. In fact, the k-means classification into $L_3$ and $L_2$ peaks worked very well. Another observation is that sLCNN-7 does not predict peaks where no peaks are expected, which is evident by the localization of KST in parts of the spectrum where $\mu_i$ and $\sigma_i$ are not vanishing. This also means large $\Delta_{i,j}$ are more likely to occur in regions of a spectrum for which spectral intensity is expected. Hence, the intensity distribution of large KSTs is somewhat proportional to the mean (or standard deviation) of the test spectra. Significant peak mismatches can occur and reach more than 50% of the intensity of the largest peak in $\mu_i$. Fortunately, the number of such cases is very small. These misfits are generally very close to the largest peak of $\mu_i$.

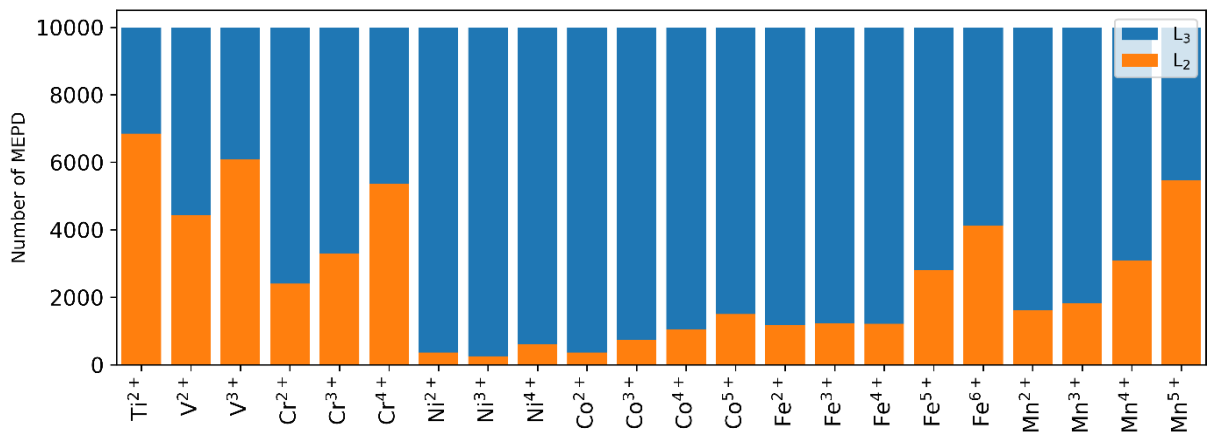

*Figure 6. Ratio between the number of KST in MEPD belonging to $L_3$ and $L_2$ peaks.*

It follows to question if an accumulation of KSTs may appear in one of the branches; that is to say, does the $L_3/L_2$ resolved MEPD correlate with the branching ratio to a similar degree that the maximal KST evolve near the largest peak in $\mu_i$. In this regard, Figure 6 provides a somewhat more characteristic description. In general, more KSTs belong to $L_3$ than $L_2$. Exceptions are Ti, V, $Cr^{4+}$, $Fe^{6+}$ and $Mn^{5+}$ which have a more equal ratio between KST in $L_3$ and $L_2$. Elements with small spin-orbit splitting in their semicore states $p_{\frac{1}{2}}$ and $p_{\frac{3}{2}}$, such as Ti, as well as cases for which the latter ratio is closer to 1, have a more equal distribution of KST among $L_2$ and $L_3$. With an increasing oxidation state, the KSTs shift to $L_2$ features. This means that some ions, e.g., $Cr^{2+}$, $Fe^{2+}$ and $Mn^{2+}$ have more robust individual peak predictions in $L_2$ than in $L_3$, whereby the $r_{MEPD}$ resembles to some extend the non-trivial branching ratio in $2p$ XAS.

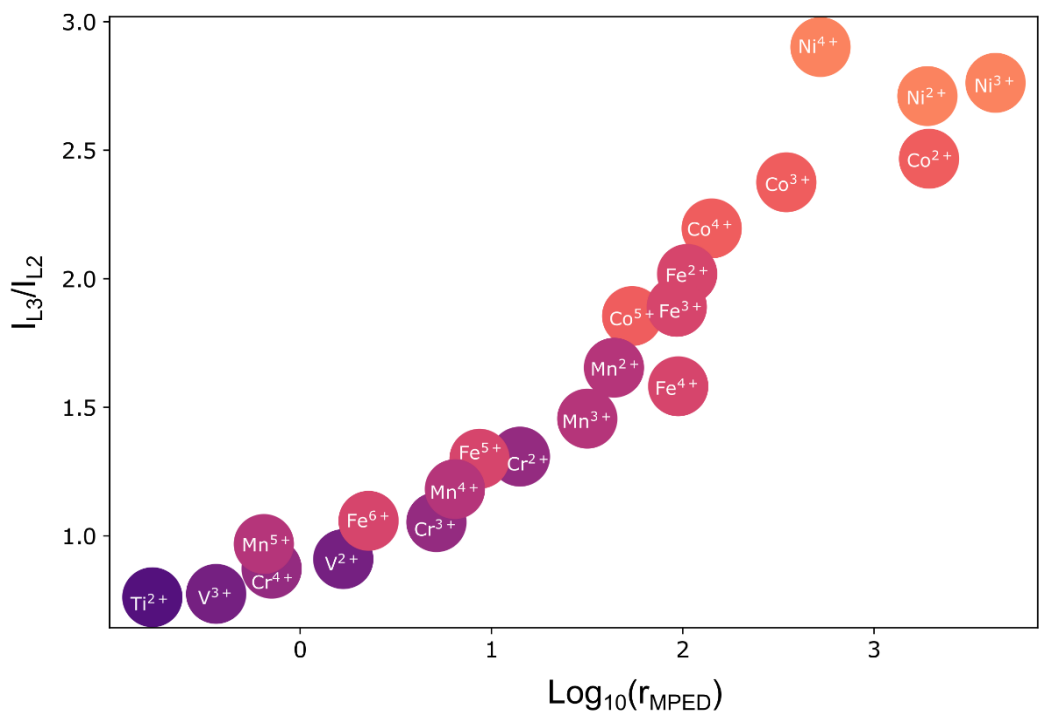

*Figure 7. Relation between the branching ratio of $L_3$ and $L_2$ in μ to the ratio of MEPD for $L_3$ and $L_2$ cases. Shading represents elements. Oxidation states are indicated.*

As noted above, the non-trivial branching ratio between L₃ and L₂ features is somewhat typical for ions in $2p$ XAS. In Figure 7, the ratio of a KST belonging to L₃ or L₂ ($r_{MPED}$) is related to the intensity ratio of L₃ and L₂ features in the average of all test spectra ($I_{L_3}/I_{L_2}$) for all tested ions. Note the figure reports the logarithm of $r_{MEPD}$. A clear trend is recognizable. With the increasing nuclear charge, $I_{L3}/I_{L2}$ as well as $r_{MEPD}$ increase confirming the correlation in MEPD concerning the intensity distribution in $2p$ XAS spectra.

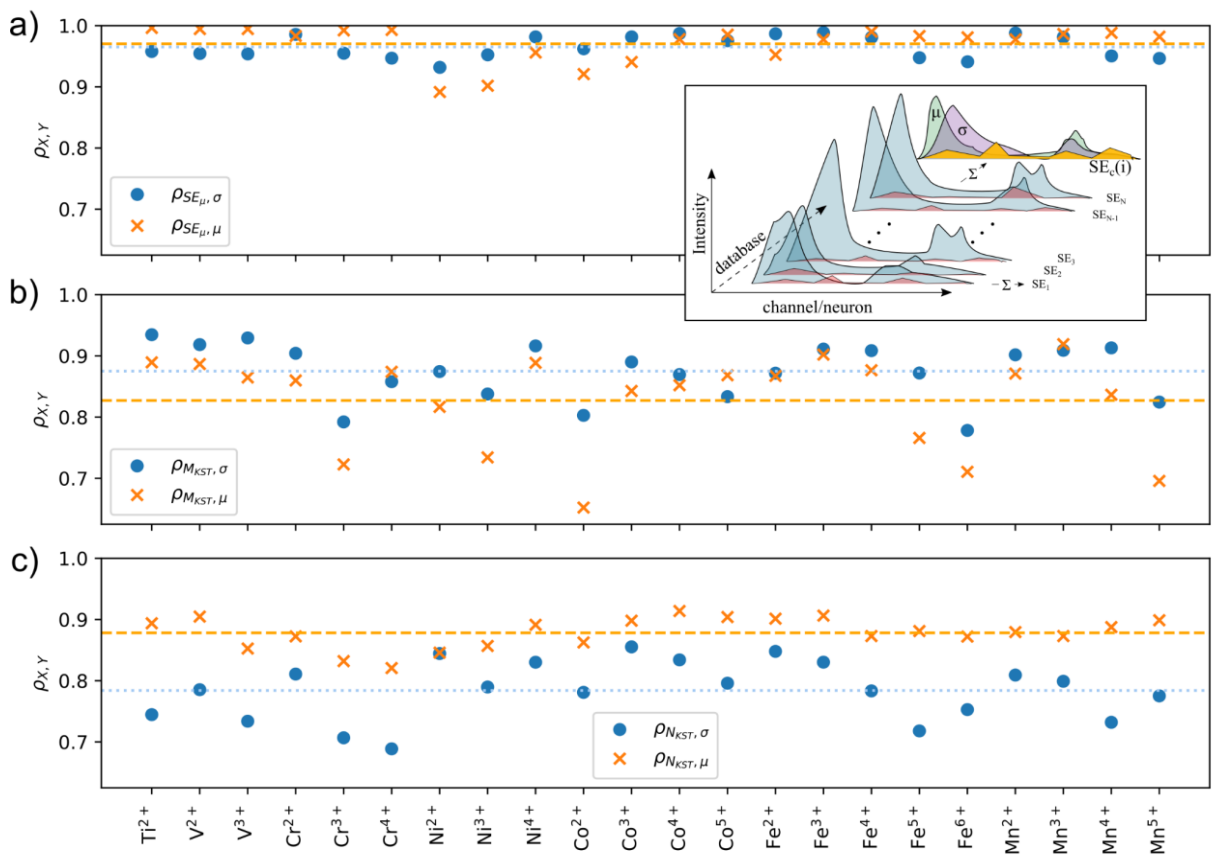

Figure 8. Pearson correlation coefficients ρ between mean ($\mu/m$) $SE_c$ and a) the channel resolved mean and standard deviation test spectra. The difference between SE, μ, σ and $SE_c$ are illustrated in the insert in which blue shaded spectra represent test data, red: differences between the latter and recomputed spectra, green: the average of all test spectra, purple: the standard deviation of the test spectra, and yellow: the channel resolved SE. b) Pearson correlation coefficient ρ between (top) max KST and the average/standard deviation (σ or μ) for all test spectra and c) the number of KTS per channel and σ or μ. Dashed and dotted lines present the averaged Pearson correlation coefficients for σ and μ, respectively.

Figure 8a reports the Pearson correlation $\rho$ between the mean of the test spectra resolved for each channel and the channel resolved mean values of the $SE$, i.e. $SE_\mu$, as $\rho_{SE_\mu,\mu}$. Also, the correlation to the standard deviation $\sigma$ is given as $\rho_{SE_\mu,\sigma}$. The insert illustrates the difference between $SE$ and $SE_c$ as well as $\mu$ and $\sigma$. Figure 8b and 8c show the Pearson correlation between the max. KSTs ($M_{KST}$) of each channel (or neuron) and the mean/standard deviation of the test spectra as well as for the number of KST values ($N_{KST}$), respectively. The difference between $M_{KST}$ and $N_{KST}$ can be taken as what is the maximal peak misfit and what is the likelihood for a channel to report a KST.

There is very strong correlation between $SE_\mu$ and the mean and standard deviation of the test spectra ($\sigma$ and $\mu$, respectively). The correlation factors are between 0.87 and reach almost 1 for some ions. $M_{KST}$ shows correlation values to $\sigma$ and $\mu$ between 0.65 and 0.96 with averages of about 0.9 where those of $\sigma$ are generally larger than for those of $\mu$. In contrast, $\rho_{N_{KST}}$ correlates stronger with the $\mu$ than with $\sigma$, however, the correlations are less with maximal values of 0.91 and averages of ca. 0.9 and 0.8, respectively. It can be summarized that the max. and std. dev. of $SE_c$ can be estimated from the std. dev. of the test spectra. Peak misfits (KSTs) are somewhat more random in this regard. The likelihood of KST increases with $\mu$ but the largest KST are seen in regions of spectra with large $\sigma$. This information could be used to estimate predictions of not tested spectra.

## High symmetry cases

The sLCNN-7 reliance testing shown above was performed for low symmetry case $D_{4h}$. Applications in the interpolation regime, that is within the present parameter ranges for the selected ions, work as expected well and are also not surprising given the ANN can, in principle, represent any real values function with arbitrary accuracy [93]. Any failure, assuming the underlying relation within the data is representable by a continuous real-valued function, would naturally be either due to insufficient computational resources, erratic or unsuitable training data or limitations in the ANN designs. In this regard, the application to crystal-field symmetries outside the training scope is somewhat more challenging and gives access to potential testing of weaknesses of the ANN-based method.

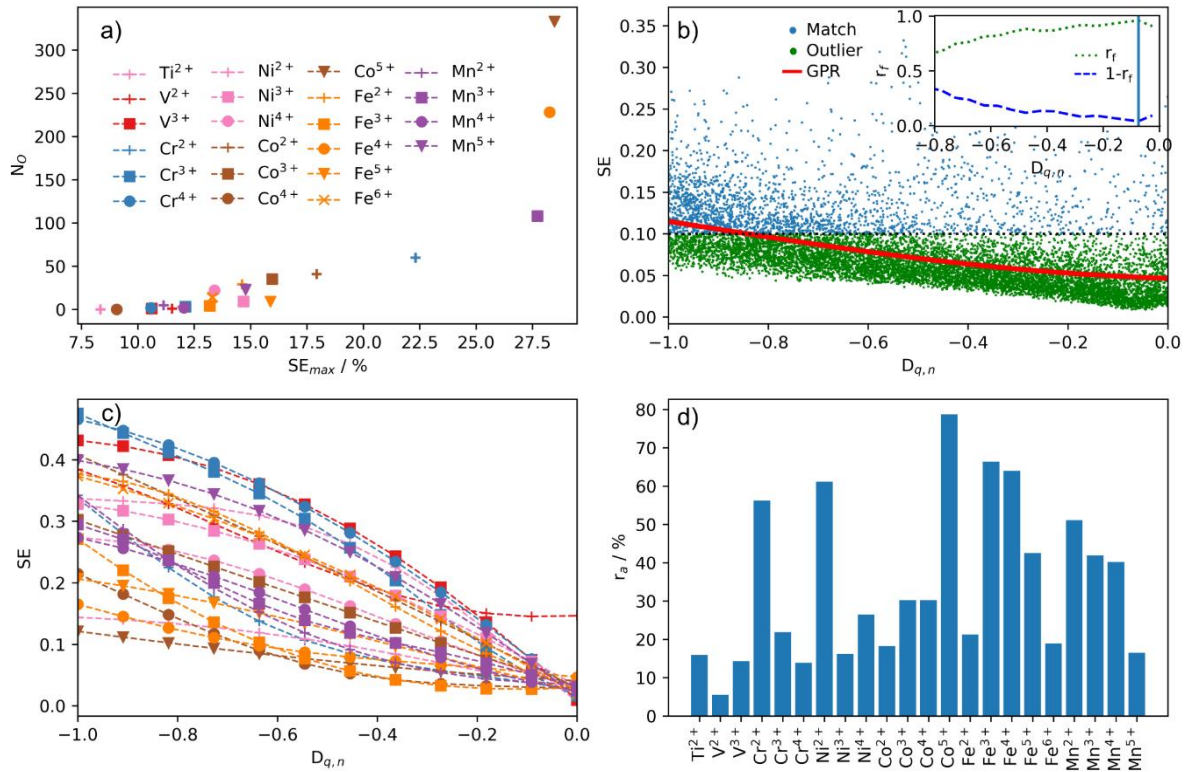

*Figure 9. Overview of high symmetry test cases. a) Results of $O_h$ test with maximal SE ($SE_{max}$) of each of the ions plotted against the number of outliers ($N_O$). b) Example of SE distribution of 10000 test spectra (match are green dots, and outliers are blue dots) for $Co^{5+}$ in $C_{4v}$ symmetric ligand field. The insert shows the $r_f$ functions and the point at with 95% of test spectra are determined accurately, i.e. matches (vertical line). c) GPR fits of $SE(D_{q,n})$ for all tested ions in $C_{4v}$ ligand fields, the curve labeling is following figure a). d) Ratio of accurately determined matches and outliers for $C_{4v}$ cases. Note that $D_{q,n}$ refers to the box-normalized $D_q$ values as it is passed to the ANNs.*

An additional performance test can be formulated for the trained ANNs concerning spectra of ions with high-symmetric ligand arrangements. This includes $O_h$ and $C_{4v}$ symmetries. Note that the former still corresponds to an interpolation because the parameter space was reduced by restricting $D_s = 0$ and $D_t = 0$. Notably, the latter two parameter choices are equal with boundaries of initial $D_{4h}$ training data intervals. With this restriction, the test performance of ANN trained on $D_{4h}$ data is not directly deducible from $D_{4h}$ test data because the random nature of parameter sampling in the database generation does not ensure that the boundaries of the selected intervals were included. This applies to training, evaluation and testing.

In contrast, the latter case ($C_{4v}$ symmetric ligand field) is formally an extrapolation concerning the Ballhausen parameters because it assumes negative values of $D_q$, here in particular: $D_q = [-0.2; 0.0]$ eV while $D_s = 0$ and $D_t = 0$. The remaining parameter ranges are unchanged. In fact, only one of the parameters ($D_q$) is in the extrapolation region while the others are still within the interpolation range. Then, it is an open question of how well the ANNs perform, or in other words, how impactful is $D_q$ in the extrapolation region on the ANNs' performance. It is worth noting at this point that extrapolation is generally a weakness of ANN.

For these tests, the $D_{4h}$-trained ANNs were used. For each ion, 10000 spectra for each of the two high symmetry cases were generated with the modified parameter ranges mentioned above. The used parameter sets were passed to the $D_{4h}$-trained ANNs and the predicted spectra were compared to the spectra obtained from the model Hamiltonian approach described above.

The $O_h$ results are summarized in Figure 9a which shows the correlation between the largest $SE$ ($SE_{max}$) and the number of cases in the test data for which the SE is larger than 10%. The number of outliers can be much larger than in the $D_{4h}$ tests. Overall, the accuracy of $O_h$ spectra is similar to those of $D_{4h}$ cases. The maximal SEs are between 8.3 and 28.5%. For most ions, maximal SE are close to 10% and their number of cases where predicted spectra's SE is larger than 10% are very few (on average 43 out of 10000). For $Mn^{3+}$, $Fe^{4+}$, and $Co^{5+}$, $SE_{max}$s are considerably larger with 27.7, 28.3 and 28.5 %, respectively. For these cases, the number of outliers is significantly increased, too. The $Co^{5+}$ reported the largest number of 333, while $Fe^{4+}$ and $Mn^{3+}$ have 228 and 108 outliers, respectively. Note that the next largest SE was seen in $Cr^{2+}$ (22.3%), but the number of outliers was only 60 and followed the trend between SE and the number of outliers seen for the other ions. Interestingly, these four ions have four d-electrons in the initial state. In the final state, they have an additional electron in $3d$ that is promoted from a 2p orbital giving them a large combinatorial complexity of electronic configurations. For the other ions, the increase of SE follows roughly the increasing combinatorial complexity of degeneracy of a simple 3d$^n$ configuration given by $\binom{10}{n} = \frac{10!}{(10-n)!n!}$. Of course, the degeneracy between d-states is partly lifted in $O_h$ systems. Then, the split between e$_g$ and t$_{2g}$ states increases with increasing values of $D_q$. For small $D_q$, however, the energy splitting might not be sufficient to result in spectral changes that were learned by the ANN. In combination with a huge amount of possible transitions, this could lead to large SEs, particularly for ions with four d-electrons in the initial state. Furthermore, transitions between initial and final states must obey dipole selection rules that influence the spectral shape significantly. This causes further complications that could be the source of increased SEs for $O_h$ and $C_{4v}$ systems.

For $C_{4v}$ systems, the relevance of $D_q$ associated with extrapolations by the ANN when $D_q$ has negative values is reflected in Pearson correlation coefficients in the magnitude of the order of 80%, which is the largest among all parameters for $C_{4v}$ ligand cases. This means that the d-level splitting in the extrapolation range has the largest impact on SE. Because a more significant number of outliers, i.e. failure of the ANN, is expected for $C_{4v}$ ligand symmetries, due to the extrapolation characteristics for the here presented ANN, the analysis was expanded by a non-parametric Gaussian process regression for the $SE_{C4}(D_q)$ curve. Additionally, a function $r_f$ is introduced that represents the ratio between successful and failed recomputed spectra (measured on the SE being smaller or larger than 10%, respectively) in intervals of $\Delta D_q = 0.05$ eV. For the latter function, a ratio of 95% (i.e. 95% of all spectra have SEs of less than 10%) serves as a marker to divide $D_q$ into regions that reliable recompute spectra in $C_{4v}$ symmetry and those that do not. The results for Co$^{5+}$, as an example, are given in Figure 9b. Note that the ratio of outliers for the O$_h$ cases can be notably larger than for the original training.

The GPR estimates of SEs for all ions are compared in Figure 9c and Figure 9d gives an overview of the $r_a$'s.

The ANNs perform only in a relatively small region outside the training parameter ranges for most ions with sufficient accuracy. Because the ranges of $D_q$ are of equal size, but they have opposite directions in the $D_{4h}$ and $C_{4v}$ databases, their d-level splitting strength can be compared. The maximal SE values for $C_{4v}$ are much larger than for the other symmetries due to the extrapolation failure at large $|D_q|$. For most cases, accurate extrapolation is only possible with the first 2.5% of the tested $D_q$ range, i.e, between -0.005 and 0 eV. In some cases, this can reach more than 10% or larger ($V^{2+}$, $Co^{5+}$, $Fe^{2+}$, $Fe^{3+}$, $Fe^{5+}$, $Mn^{2+}$ and $Mn^{4+}$). This limited range often causes small $r_a$ of less than 20%. The $Co^{4+}$ is an exception for which $D_q$ close to -0.1 eV still gives accurate spectra with 95% likelihood. Similarly surprising is the ANN for $Co^{5+}$, which yielded a $r_a$ of more than 78%, while the average $r_a$ of all ions is only 34% with a median at 28.3%. Notably, $Co^{4+}$ and $Co^{5+}$ have rather flat GPR curves as well as flat $r_a$ functions compared to other ions. This means the SEs are more accurate over a large range of $D_q$ values than for other ions. In fact, GPR fits present the most likely estimate for SE at given values of $D_q$. Large dependence on the ion is also seen in shape as and the maximal values for the fits. They range between 0.11 ($Co^{5+}$) and 0.48 ($Cr^{3+}$) at $D_q$ of $-0.2$ eV.

It was also noted that some SE values were larger than 50% (i.e., a complete mismatch of two spectra normalized by their superposition), which is beyond the physically meaningful limit. sLCNN-7 reporting negative intensities caused this for some channels in the extrapolation regime. This behavior results from the linear activation function used in the output layer and is an indication that the model did not capture the underlying physics sufficiently well for $C_{4v}$ systems.

Recently Roest *et al.* suggested using the ReLU function in the output layer to present such unphysical results [102]. However, unphysical outcomes in $C_{4v}$ of the ANN were not observed for all ions. It could be argued that the ANN training was more suitable in those cases resulting in better learning of the underlying physics within the given training data. In contrast, the same level of progress was simply not achieved for those reporting unphysical conditions. Then, it could be speculated that further effort to improve training could limit unphysical behaviour of the ANNs. Also, the extrapolation regions where this was encountered are known to be a weakness of ANN. The well-performing cases rather are exceptions.

Following the above-presented application, it is obvious that trained ANNs struggle with extrapolation, e.g., almost 95 % outliers for $V^{2+}$. Even in the cases for interpolation tasks of $O_h$, the number of outliers slightly increases compared to $D_{4h}$ situations. The reported results indicate that individual analysis of ANN performance is required. Generally, it cannot be assumed that a type of ANN designed and trained for an element and ion can automatically achieve the same accuracy for other systems even if ANNs are individually trained. This applies especially to applications that require extrapolations of one or more input features.

## Comparison to experimental results

Lastly, the trained ANNs are tested using experimental results. For this purpose, a selection of $2p$ XAS of four well-known materials includes two monoxides and two metal-phthalocyanines (MPc): CoO, FeO, CoPc and FePc. CoO and FeO cover tests for $O_h$ ligand fields. Spectra of Co- and Fe-phthalocyanine (CoPc and FePc, respectively) evaluate the ANNs for $D_{4h}$ cases. In an earlier study, the inversion by ANN of $2p$ XAS spectra back to a parametrized form was presented [80]. Here, the parameter values from that study are given in Table 4 and used to predict the spectra of the four materials with sLCNN-7. Additionally, the table reports the SE between the predicted spectra and the experimental ($SE_{exp}$) as well as the MHA obtained spectra

($SE_{MHA}$). Figure 10 compares the experimental, predicted and the MHA spectra. The box normalization was applied to all experimental spectra.

The agreement between MHA spectra and sLCNN-7 prediction for all spectra is at least good. The $SE_{MHA}$ values are not larger than 3% but for FeO for which a missing feature at 710 eV and small deviation in $L_2$ lead to $SE_{MHA}$ of 5.5%. The seen differences are within the expectation given the above analysis. For instance, the largest peak misfit is seen in $L_3$ in $Fe^{2+}$ as indicated by its $r_{MPED}$, while other spectral differences are rather small. Given the above analysis, it is expected that the method performs slightly less accurate for $O_h$ than for $D_{4h}$. The $Fe^{2+}$ cases confirm this.

Larger differences are, as expected, seen in comparison to the experimental spectra. The MHA as well as the sLCNN-7 suffer from the same limitations because the sLCNN-7s' training is based on the results of MHA. This means that any disability to account for spectral features is inherited by sLCNN-7. Such contributions to a spectrum could come from, for instance, defects, multipole contributions, phonons or simply transitions into states not described by the MHA, e.g., 3s or condition bands. Hence, differences seen between experiments and simulated/predicted spectra must be evaluated on a different scale. In this regard, an absolute measure of SE might be less applicable, but its minimization might find practical applications. CoO has the largest $SE_{exp}$ of more than 25% caused by underestimated tail regions in $L_3$ and $L_2$ as well as an overestimated peak at about 779.5 eV. A SE of more than 25% is more acceptable in this case because the method yielded similar agreement between theory and experiment as in previous studies [4,103,104], while a SE of 25% indicates, in general, a mismatch. The ANNs obtain better agreement with SEs of about 20% for CoPc and FePc. For these systems, the overall spectral shape follows the experimental while a few peaks are underestimated. The $SE_{exp}$ is significantly smaller with 9.1% for FeO. The good agreement of the latter case exemplifies the potential for particular applications.

Table 4. Parameters of previously published 2p XAS simulations from Ref. [80] and their SE between sLCNN-7 and MHA. SE with regard to the experimental spectrum ($SE_{exp}$) and to the MHA ($SE_{MHA}$) are given as well.

| Mat. | $Co^{2+}$ | $Co^{2+}$ | $Fe^{2+}$ | $Fe^{2+}$ |
|---|---|---|---|---|
|  | CoPc | CoO | FePc | FeO |
| Sym. | $D_{4h}$ | $O_h$ | $D_{4h}$ | $O_h$ |
| $D_q$ /eV | 0.0043 | 0.0083 | 0.13 | 0.022 |
| $D_s$ /eV | 1.0 | 0.0 | 0.35 | 0.0 |
| $D_t$ /eV | -0.13 | 0.0 | 0.037 | 0.0 |
| $S_1$ | 0.91 | 0.82 | 0.62 | 0.9 |
| $S_2$ | 0.79 | 0.81 | 0.86 | 0.92 |
| T/K | 708 | 303 | 8 | 848 |
| $\Gamma_1$ /eV | 0.44 | 0.68 | 0.59 | 0.77 |
| $\Gamma_2$ /eV | 0.75 | 0.73 | 0.76 | 1.03 |
| $\sigma_G$ /eV | 0.2 | 0.2 | 0.1 | 0.1 |
| $SE_{exp}$ /% | 17.1 | 26.0 | 19.8 | 9.11 |
| $SE_{MHA}$ /% | 2.76 | 1.76 | 1.99 | 5.49 |

Overall, the achieved performance of sLCNN-7 for 2p XAS spectra prediction is encouraging. To further enhance the accuracy and reliability of the method, other neural network architecture, for instance, the deep tensor neural network used by Rankine *et al.*[71] to predict excitation spectra of molecules, could be explored for their suitability for 2p XAS prediction, although modifications based on the limited number of input feature may be required.

The above analysis and application examples of the developed ANN-based method demonstrate the capabilities of the approach and sufficient confidence in the method was established to consider it a reliable alternative of computing 2p XAS spectra to first-principles or model Hamiltonian methods. In this regard, a combination of different statistical measures appears beneficial to evaluate the performance of ANNs. An essential aspect therein is the possible failure of the method and conditions that lead to them. This can also allow future studies compare to ANN performance. It also is worth mentioning that, in principle, any other working methods based on first principles approaches simulating 2*p* XAS spectra could replace the employed MHA generating the reference spectra. As such, neither the success nor the failure of the ANN is surprising due to its foundation and statistical nature.

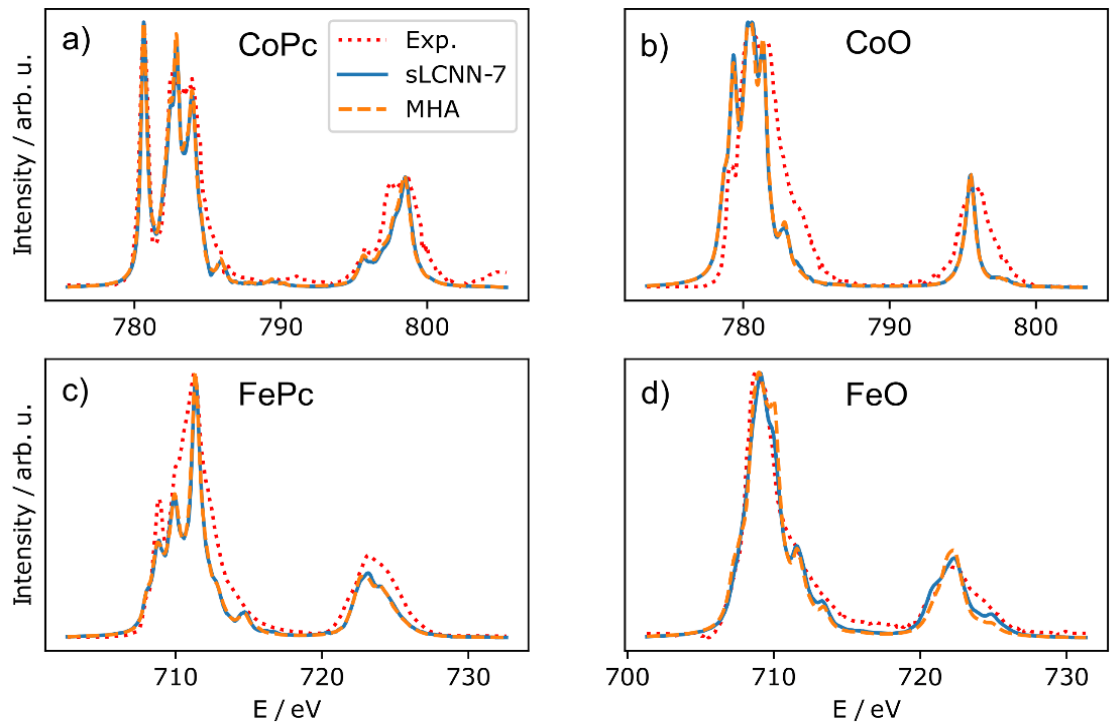

*Figure 10. Comparison between experimental, predicted and MHA obtained spectra for a) CoPc, b) CoO, c) FePc and d) FeO. Experimental data were digitized from Ref. [101] (CoPc), [105] (CoO), [7](FePc) and [100] (FeO). Background contributions in the experimental spectra were manually removed [80].*

With the recent rise of machine learning techniques and their widely available implementations, questions arise on how theoretical x-ray spectroscopy and consequently experimental advances in this field can benefit from them. Since several simulation approaches already provide excellent reproduction of features in experiments besides the recently demonstrates capabilities of ANN to invert the simulations and experimental spectra [80], the performance target for ANNs to predict spectra should be on speed while maintaining accuracy. Then, the ML contribution to the field could be expanded to high-throughput workflows through computational performance benefits, that might find application in advanced XAS imaging approaches [106] or in situations where materials structure has a complexity beyond that of, e.g., simple transition metal monoxides.

There are also advantages concerning high data volume. For example, the prediction of 10000 spectra for each of the ions took only a few seconds, while the MHA computation of them took depending on the ion days to weeks on a modern multicore CPU. However, a direct comparison of computation speed is difficult because the ANNs use a GPU and not the same CPU. Still, this difference shows that in high-throughput setups, e.g., 2$p$ XAS spectromicroscopy in which each pixel contains a full spectrum and images of $100 \times 100$ pixels equal 10000 spectra, could be handled time efficiently by the presented ANNs.

## Concluding Remarks

In summary, an ANN approach predicting 2$p$ XAS spectra for 22 light transition metal ions is presented. The method takes a set of parameters describing atomic and ligand properties as well as information about experimental conditions, including temperature and convolutions, to predict spectra. There is a notable dependence on ANN architectures type to capture the complexity of the 2$p$ XAS process. Also, the training dataset size is critical for successful ANN training. Either too small datasets or too simple ANN designs can prevent successful model training. High accuracy was achieved and, in general, 99% of predicted spectra are highly accurate. Errors in peak intensity prediction roughly correlate with $L_3/L_2$ peak shapes and their branching ratios that are characteristic of each transition metal ion. Also, a correlation between uncertainties of ANN predictions and the overall complexity of spectra was observed. This could provide access to outlier detection and confidence estimates for untested cases. Although none of the tested ANN architectures has the same level of accuracy as the exact model Hamiltonian approach that provided the data for the training, the current results point out the current capabilities of ANN concerning 2p XAS prediction. This even applies partly to applications of the ANN in extrapolation conditions. Moreover, its ability to reproduce experimental spectra was demonstrated and achieved the same degree of agreement as the MHA. By comparison with experimental spectra, it was shown that the solution of the direct problem by ANN is compatible with the solutions of the inverse problem from a previous study. Still, the presented ANN approach for simulating 2p XAS spectra is potentially more suitable for high-throughput setups, as they could be needed for 2p XAS based spectromicroscopy, given its computational efficiency, confirmed accuracy and reliability.

# Acknowledgments

The presented research was financially supported by the Ministry of Science and Technology Taiwan (MOST) under Grant no. MOST109-2112-M-110-009. J.L. thanks the National Center for High-performance Computing (NCHC) of National Applied Research Laboratories (NARLabs) in Taiwan for providing computational and storage resources.